\documentclass[twocolumn,english]{IEEEtran}
\usepackage[T1]{fontenc}
\usepackage[latin9]{inputenc}
\usepackage{xcolor}
\usepackage{amsbsy}
\usepackage{stackrel}
\usepackage{graphicx}
\usepackage{babel}

\usepackage{fancyhdr}

\begin{document}	
\title{Exceptional Points of Degeneracy Directly Induced by Space-Time Modulation
of a Single Transmission Line}
\author{\author{Kasra Rouhi, Hamidreza Kazemi, Alexander Figotin, and Filippo Capolino, \it{Fellow, IEEE} \thanks{K. Rouhi, H. Kazemi, and F. Capolino are with Department of Electrical Engineering and Computer Science, University of California, Irvine, CA 92697, USA.}
\thanks{A. Figotin is with Department of Mathematics, University of California, Irvine, CA 92697, USA.}
\thanks{(e-mail: {kasra.rouhi, hkazemiv, afigotin, f.capolino}@uci.edu)}}}
\maketitle

\thispagestyle{fancy}

\begin{abstract}
We demonstrate how exceptional points of degeneracy (EPDs) are induced
in a single transmission line (TL) directly by applying periodic space-time
modulation to the per-unit-length distributed capacitance. In such
space-time modulated (STM)-TL, two eigenmodes coalesce into a single
degenerate one, in their eigenvalues (wavenumbers) and eigenvectors
(voltage-current states) when the system approaches the EPD condition.
The EPD condition is achieved by tuning a parameter in the space-time
modulation, such as spatial or temporal modulation frequency, or the
modulation depth. We unequivocally demonstrate the occurrence of the
EPD by showing that the bifurcation of the wavenumber around the EPD
is described by the Puiseux fractional power series expansion. We
show that the first order expansion is sufficient to approximate well
the dispersion diagram, and how this \textquotedblleft exceptional''
sensitivity of an STM-TL to tiny changes of any TL or modulation parameter
enables a possible application as a highly sensitive TL sensor when
operating at an EPD.
\end{abstract}


\section{Introduction}

Recent advancements in EPD concepts have attracted a surge of interests
due to their potential benefits in various electromagnetic applications.
An EPD is a point in parameter space of a system at which multiple
eigenmodes coalesce in both their eigenvalues and eigenvectors. The
concept of EPD has been studied in lossless, spatially \cite{Figotin2003Oblique,Nada2017Theory,Othman2016GiantGain}
or temporally \cite{Kazemi2019Exceptional} periodic structures, and
in systems with loss and/or gain under parity-time symmetry \cite{Bender1998Real,Heiss2004Exceptional,Schindler2011Experimental,Hodaei2014Parity}.
Since the characterizing feature of an exceptional point is the strong
full degeneracy of at least two eigenmodes, as implied in \cite{Berry2004Physics},
we stress the importance of referring to it as a \textquotedblleft degeneracy\textquotedblright ,
hence of including the D in EPD. In essence, an EPD is obtained when
the system matrix is similar to a matrix that comprises a non-trivial
Jordan block \cite{Figotin2003Oblique,Figotin2005Gigantic,Othman2017Theory,Abdelshafy2019Exceptional},
here however the formulation leads to a matrix of infinite dimensions
and therefore we assess the occurrence of the EPD by invoking the
Puiseux fractional power expansion series \cite{Kato1995Perturbation}
to describe the bifurcation of the dispersion diagram at the EPD.
There are several features associated with the development of EPDs,
which lead to applications, such as active systems gain enhancement
in waveguides \cite{Othman2016Theory,Abdelshafy2018Electron,Veysi2018Degenerate},
and enhanced sensing \cite{Wiersig2016Sensors,Chen2018Generalized,Kazemi2019Experimental,Kazemi2019Ultra}.
\begin{figure}
\begin{centering}
\includegraphics[width=2.8in]{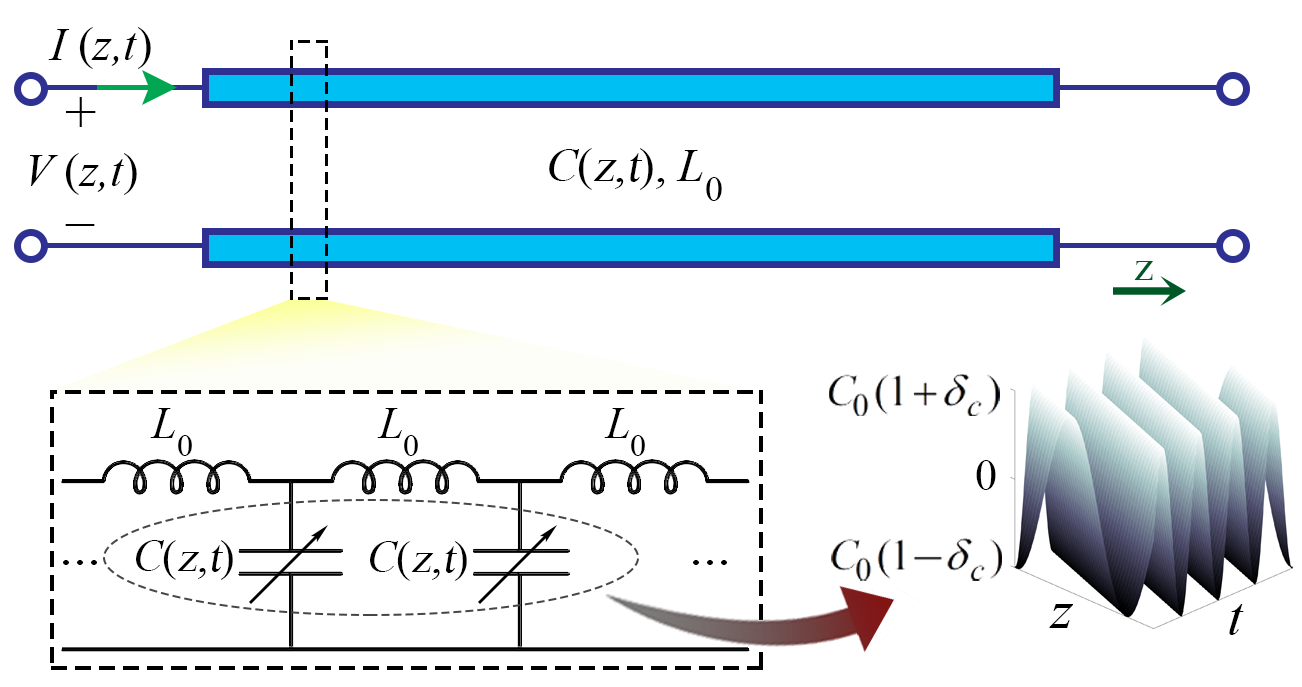}
\par\end{centering}
\caption{Schematic illustration of a single TL with space-time modulation of
the distributed capacitance. We also show the equivalent TL circuit
with the per-unit-length inductor and modulated capacitor.}
\label{fig: schem_STVTL}
\end{figure}

Researchers have been studying how to incorporate time-variation of
parameters into electromagnetic systems with the goal of adding new
degrees of freedom in wave manipulation. In their pioneering work,
Cassedy and Oliner studied the dispersion characteristics of wave
propagation in a medium with dielectric constant modulated as a traveling-wave
with sinusoidal form \cite{Cassedy1963Dispersion,Cassedy1967Dispersion}.
Then, Elachi studied electromagnetic wave propagation and the wave
vector diagram in general space-time periodic materials for different
wave polarization \cite{Elachi1972Electromagnetic}. In \cite{Zurita-Sanchez2009Reflection,MartinezRomero2016Temporal},
authors analyzed the reflection and transmission of an incident wave
onto a time-periodic dielectric slab and generalize the concept of
temporal photonic crystals to periodic modulation of the permeability
and permittivity. In \cite{Estep2016Magnetless}, magnetless nonreciprocity
was demonstrated in spatiotemporally modulated coupled-resonator networks.
Also, Taravati et al. proposed a mixer-duplexer-antenna leaky-wave
system based on periodic space-time modulation \cite{Taravati2017Mixer}.
Recently, in \cite{Rajabalipanah2020Reprogrammable}, space-time modulation
was employed to control both phase and amplitude tunability in a metasurface,
whereas traditionally metasurfaces have phase-only control on reflection/transmission.
Several other papers have been published in recent years on time/space-time
modulation to generate nonreciprocity in electromagnetic structures
as \cite{Qin2014Nonreciprocal,Lurie2016Energy,Reiskarimian2018Analysis,Wu2019Magnetic,Wu2019Isolating}.
In all these works, the concept of EPD in such modulated structures
was not studied.

Here we leverage on the two concepts of space-time modulation and
EPD and develop a general scheme to realize EPDs in space-time periodic,
single TLs that can be used as a sensor or leaky wave antenna with
ultra-high sensitivity. We investigate the occurrence of EPDs in a
single TL when the per-unit-length capacitance is modulated in space
and time, and we demonstrate its occurrence by showing that the bifurcation
of the dispersion diagram around the EPD is well approximated by the
Puiseux fractional power series expansion. This EPD-related fractional
expansion is also used to explain the extreme sensitivity of the wavenumber
to perturbation of system parameters. 
\begin{figure}
\begin{centering}
\includegraphics[width=2.8in]{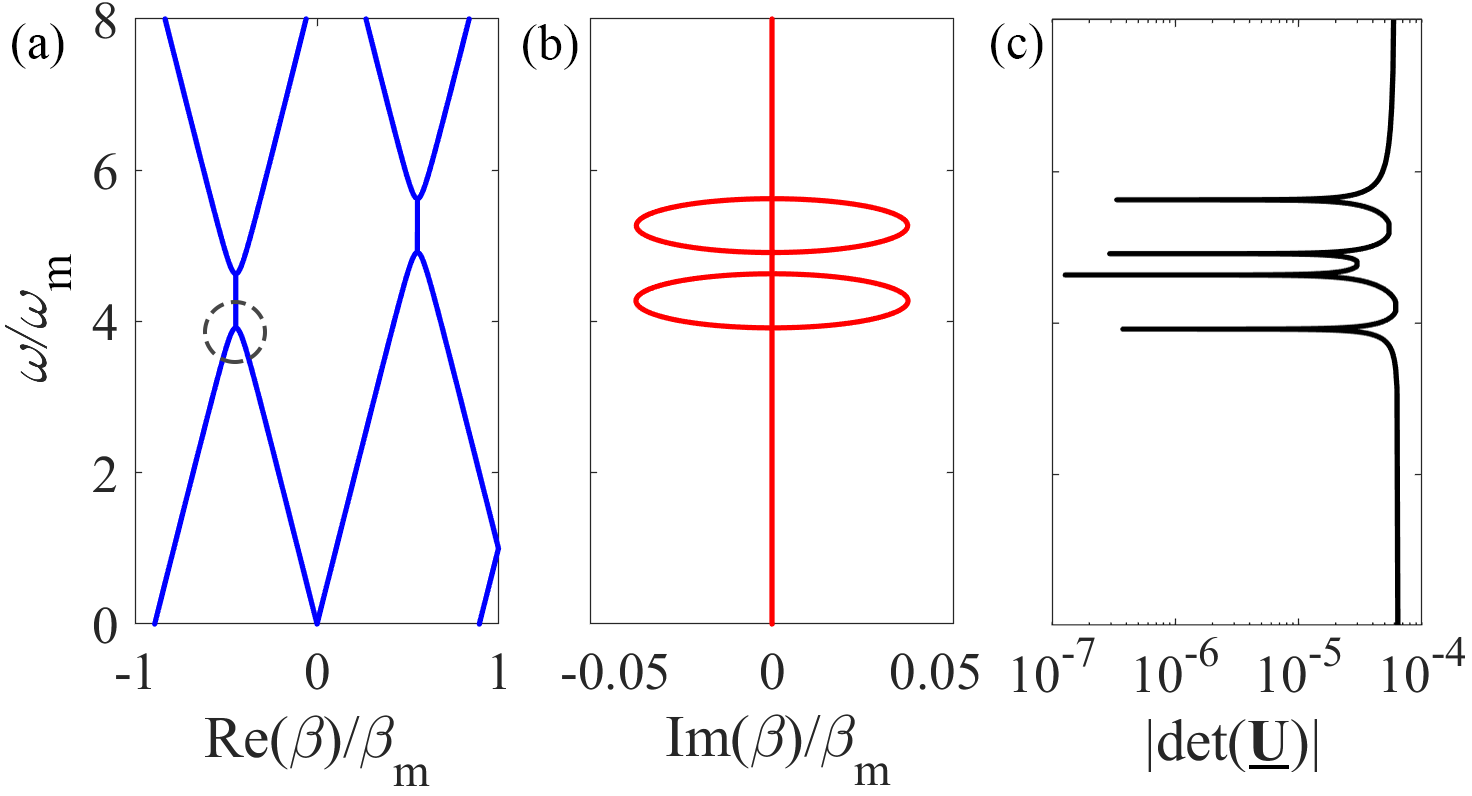}
\par\end{centering}
\caption{Dispersion diagram of the STM-TL with 2nd order EPDs. (a) Real part
of the wavenumber of first harmonics and (b) the corresponding imaginary
parts, and (c) plot of $\left|\mathrm{\mathrm{det}}(\underline{\mathbf{U}})\right|$
versus frequency. The similarity transformation matrix $\mathbf{\underline{U}}$
contains eigenvectors, therefore the vanishing of $\left|\mathrm{\mathrm{det}}(\underline{\mathbf{U}})\right|$
is necessary at an EPD.}

\label{fig: Dispersion}
\end{figure}

\section{Degeneracies in a Uniform Single ST-MTL}

Degeneracies in wave propagation in an infinitely long TL is examined
when the per-unit-length capacitance is modulated in both space and
time. We employ the formalism and description of a linear TL shown
in \cite{Othman2016Low}. A schematic representation of an STM-TL
is shown in Fig. \ref{fig: schem_STVTL}, where only the per-unit-length
capacitance is space-time varying, while the per-unit-length inductance
is constant throughout the TL. Without loss of generality we assume
sinusoidal space-time variation; however, EPDs can be induced also
by other forms of periodic space-time variation. The distributed per-unit-length
space-time varying capacitance is given by

\begin{equation}
C(z,t)=C_{\mathrm{0}}+C_{m}(t)=C_{0}(1+\delta_{\mathrm{c}}\cos(\omega_{\mathrm{m}}t-\beta_{\mathrm{m}}z)),\label{eq: C_STV}
\end{equation}
where $C_{\mathrm{0}}$ is the space-time averaged (i.e., unmodulated)
per-unit-length capacitance, $\delta_{\mathrm{c}}$ is the modulation
depth, and $\omega_{\mathrm{m}}$ and $\beta_{\mathrm{m}}$ are the
temporal and spatial modulation frequencies, respectively. The dynamic
behavior of such a TL is captured using the Telegrapher\textquoteright s
equations, that are here represented in terms of a voltage and current
state vector, $\boldsymbol{\boldsymbol{\Psi}}(z,t)=[\begin{array}{cc}
V(z,t), & I(z,t)\end{array}]^{\mathrm{T}}$, where the superscript $\mathrm{T}$ denotes the transpose operation.
The dynamic behavior of this state vector is described by the first
order differential equations as
\begin{eqnarray}
\partial_{\mathrm{z}}\boldsymbol{\boldsymbol{\Psi}}(z,t) & = & -\partial_{\mathrm{t}}\left(\underline{\mathbf{\underline{M}}}(z,t)\boldsymbol{\boldsymbol{\Psi}}(z,t)\right),\label{eq: Telegraph_Eq}
\end{eqnarray}
where the space-time modulated $2\times2$ system matrix $\mathbf{\underline{\underline{M}}}$
is given by

\begin{equation}
\mathbf{\underline{\underline{M}}}(z,t)=\left[\begin{array}{cc}
0 & L_{\mathrm{0}}\\
C(z,t) & 0
\end{array}\right]\;.\label{eq: M_eq}
\end{equation}

We look for time-harmonic solutions, and because of the periodic nature
of the modulation the state vector eigensolution is cast into an infinite
space-time Floquet-Bloch series as

\begin{equation}
\boldsymbol{\boldsymbol{\Psi}}(z,t)=e^{j(\omega t-\beta z)}\stackrel[q=-\infty]{\infty}{\sum}\mathbf{\boldsymbol{\boldsymbol{\Psi}}}_{q}\,e^{jq(\omega_{\mathrm{m}}t-\beta_{\mathrm{m}}z)},\label{eq: state_vec_harmo}
\end{equation}
\begin{figure*}
\begin{centering}
\includegraphics[width=6.1in]{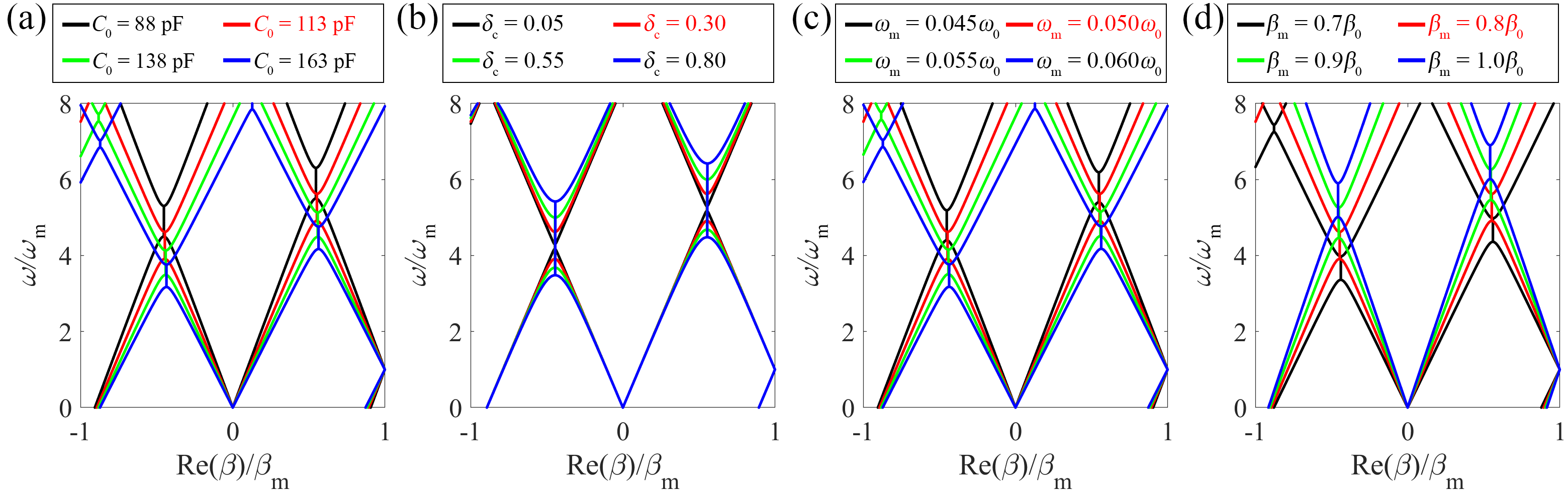}
\par\end{centering}
\caption{Dispersion diagrams of the real part of wavenumbers by changing one
single parameter at the time. The system parameters that are changed
are (a) $C_{\mathrm{0}}$, (b) $\delta_{\mathrm{c}}$, (c) $\omega_{\mathrm{m}}$,
and (d) $\beta_{\mathrm{m}}$.}

\label{fig: Sweep}
\end{figure*}
where $\beta$ and $\omega$ are the propagation wavenumber and the
angular frequency, respectively, and $\boldsymbol{\boldsymbol{\Psi}}_{q}=[\begin{array}{cc}
V_{q}, & I_{q}\end{array}]^{\mathrm{T}}$ is the complex amplitude of the $q$-th harmonic of the state vector.
We expand the space-time-varying distributed capacitance in Eq. (\ref{eq: C_STV})
in terms of its Fourier series

\begin{equation}
C(z,t)=\stackrel[s=-1]{1}{\sum}C_{s}\,e^{js(\omega_{\mathrm{m}}t-\beta_{\mathrm{m}}z)},\label{eq: Cap_harmo}
\end{equation}
where $C_{s}$ represents the amplitude of the $s$-th harmonic. Substituting
Eqs. (\ref{eq: state_vec_harmo}) and (\ref{eq: Cap_harmo}) in Eq.
(\ref{eq: Telegraph_Eq}) and taking the time and space derivatives,
the equation for each $q$-th harmonics\textquoteright{} $\mathbf{\mathbf{\mathbf{\boldsymbol{\boldsymbol{\Psi}}}}}_{q}$
is obtained as
\begin{equation}
\begin{array}{c}
\stackrel[q=-\infty]{\infty}{\sum}(\beta+q\beta_{\mathrm{m}})\mathbf{\boldsymbol{\boldsymbol{\Psi}}}_{q}\,e^{jq(\omega_{\mathrm{m}}t-\beta_{\mathrm{m}}z)}=\\
\stackrel[q=-\infty]{\infty}{\sum}\stackrel[s=-1]{1}{\sum}\left[\begin{array}{cc}
0 & (\omega+q\omega_{\mathrm{m}})L_{\mathrm{0}}\delta_{s,\mathrm{0}}\\
(\omega+(q+s)\omega_{\mathrm{m}})C_{s} & 0
\end{array}\right]\\
\mathbf{\boldsymbol{\boldsymbol{\Psi}}}_{q}\,e^{j(q+s)(\omega_{\mathrm{m}}t-\beta_{\mathrm{m}}z)},
\end{array}\label{eq: diff_eq}
\end{equation}
where $\delta_{s,\mathrm{0}}$ is the Kronecker delta. Since the exponential
functions $e^{jq(\omega_{\mathrm{m}}t-\beta_{\mathrm{m}}z)}$ form
a complete orthogonal set of functions, we balance the coefficient
of the exponential with the same $q$ index leading to

\begin{equation}
\begin{array}{c}
(\beta+q\beta_{\mathrm{m}})\mathbf{\boldsymbol{\boldsymbol{\Psi}}}_{q}=\\
\stackrel[s=-1]{1}{\sum}\left[\begin{array}{cc}
0 & (\omega+q\omega_{\mathrm{m}})L_{\mathrm{0}}\delta_{s,\mathrm{0}}\\
(\omega+q\omega_{\mathrm{m}})C_{s} & 0
\end{array}\right]\mathbf{\boldsymbol{\boldsymbol{\Psi}}}_{q-s}.
\end{array}\label{eq: diff_eq_2}
\end{equation}

Isolating the term with the wavenumber, the above equation is rearranged
as

\begin{equation}
{\color{blue}{\color{olive}\begin{array}{c}
{\color{blue}{\color{black}{\normalcolor \beta\boldsymbol{\boldsymbol{\Psi}}_{q}}}}{\color{black}=\stackrel[s=-1]{1}{\sum}\mathbf{\underline{\underline{N}}}_{q,s}\mathbf{\boldsymbol{\boldsymbol{\Psi}}}_{q-s}},\end{array}}}\label{eq: diff_eq_3}
\end{equation}
where

\begin{equation}
{\color{blue}{\color{olive}\begin{array}{c}
{\color{blue}{\color{black}{\normalcolor \mathbf{\underline{\underline{N}}}_{q,s}=\left[\begin{array}{cc}
-q\beta_{\mathrm{m}}\delta_{s,\mathrm{0}} & (\omega+q\omega_{\mathrm{m}})L_{\mathrm{0}}\delta_{s,\mathrm{0}}\\
(\omega+q\omega_{\mathrm{m}})C_{s} & -q\beta_{\mathrm{m}}\delta_{s,\mathrm{0}}
\end{array}\right]}}}\end{array}}}.\label{eq: diff_eq_N}
\end{equation}

The above equation can be cast in terms of a large block three-diagonal
matrix $\mathbf{\underline{T}}$ as
\begin{equation}
\mathbf{\underline{T}}\boldsymbol{\boldsymbol{\Psi}}=\beta\boldsymbol{\boldsymbol{\Psi}}\label{eq: T_mat}
\end{equation}
that can be sued to determine the system eigenvalues $\beta$ and
eigenvectors $\boldsymbol{\boldsymbol{\Psi}}=\left[\boldsymbol{\boldsymbol{\Psi}}_{-Q},...,\boldsymbol{\boldsymbol{\Psi}}_{0},...,\boldsymbol{\boldsymbol{\Psi}}_{Q}\right]^{\mathrm{T}}$.
A finite number $2Q+1$ of harmonics is sufficient to determine the
STM-TL wave characteristics and the occurrence of EPDs, hence the
dimension of the matrix $\mathbf{\underline{T}}$ is $2(2Q+1)\times2(2Q+1)$.
The real and imaginary parts of the wavenumber in the $\beta$-$\omega$
dispersion diagram are plotted in Figs. \ref{fig: Dispersion}(a)
and (b), respectively, for the STM-TL with parameters as follows.
As specified, we have considered the sinusoidal modulation given in
Eq. \ref{eq: C_STV} where the modulation parameters are $\delta_{\mathrm{c}}=0.3$,
$\omega_{\mathrm{m}}=0.05\omega_{\mathrm{0}}$, and $\beta_{\mathrm{m}}=0.8\beta_{\mathrm{0}}$,
where $\beta_{\mathrm{0}}=\omega_{\mathrm{0}}/c$ is the free space
propagation wavenumber at $\omega_{\mathrm{0}}/(2\pi)=10^{9}\,\mathrm{s}^{-1}$.
Moreover, the TL parameters are $L_{\mathrm{0}}=282\,\mathrm{nH/m}$
and $C_{\mathrm{0}}=113\,\mathrm{pF/m}$. Note that the modulation
frequency does not need to be comparable to the one of the radio frequency
wave. We consider $2Q+1=21$ harmonics to calculate the dispersion
diagram (we checked that a larger number provides the same result),
but we show only the first two harmonics, i.e., the real part of their
wavenumbers and the relevant imaginary parts. It is observed form
the dispersion diagram in Fig. \ref{fig: Dispersion}(a) that for
an STM-TL the band-gap locations form a tilted line, which indicates
non symmetric dispersion ($\omega(-\beta)\neq\omega(\beta)$) in such
a structure, as already pointed out in \cite{Cassedy1963Dispersion,Yu2009Complete}.
Furthermore, it is clear from this figure that the eigenvalues, i.e.,
the propagation wavenumbers of the system, are coalescing at the band
edges. To fully characterize an EPD, we have to show that the two
eigenvectors corresponding to the two coalescing eigenvalues are also
coalescing at the band edges. We define the similarity transformation
matrix as $\mathbf{\underline{U}}=[\mathrm{\mathbf{U}}_{1}\,|\,\cdots\,|\:\mathbf{U}_{2(2Q+1)}]$,
where $\mathbf{\underline{U}}_{i}$ is the eigenvector corresponding
to the $i$-th eigenvalue, and such matrix diagonalizes the system
matrix as $\mathbf{\underline{T}}=\mathbf{\underline{U}\underline{\Lambda}\underline{U}^{\mathrm{-1}}}$.
At the EPD two eigenvectors become linearly dependent (they coalesce),
therefore we verify that $\left|\mathrm{det}(\underline{\mathbf{U}})\right|$
vanishes at each EPD as a necessary condition, as shown in Fig. \ref{fig: Dispersion}(c)
\cite{Abdelshafy2019Exceptional}. Indeed, at $\omega/\omega_{\mathrm{m}}=3.911291$
we observe that two eigenvalues as well as the two associated eigenvectors
are equal to each other up to the 6 decimal digit. A sufficient condition
to assess the occurrence of an EPD without looking directly at the
eigenvectors is explained in the next section, by demonstrating that
the dispersion diagram bifurcates at the EPD following the Puiseux
fractional power expansion \cite{Kato1995Perturbation}. One may note
that it is also possible to achieve EPDs in systems with only space
modulation \cite{Abdelshafy2019Exceptional,Figotin2007Slow}, however
such systems are reciprocal, or with only time modulation \cite{Kazemi2019Exceptional}.

\section{Puiseux Fractional Power Expansion and High Sensitivity\label{sec:Sensitivity-to-perturbations}}

Extreme sensitivity to system perturbations is an intrinsic characteristic
of EPDs and this is intrinsically related to the Puiseux series \cite{Kato1995Perturbation,Moro1997Lidskii,Welters2011Explicit,Hanson2018Exceptional}
that univoquely describe the EPD occurrence. We first demonstrate
how the dispersion diagram varies by changing different system parameters,
then we show the extreme sensitivity of the wavenumber to a system
perturbation when operating at an EPD that follows the description
of the Puiseux fractional power expansion. We analyze the STM-TL wavenumbers
by varying one system parameter at the time around the value used
in the example. As a first parameter, we vary the unmodulated per-unit-length
capacitance of the TL $C_{\mathrm{0}}$ and observe its effect on
the dispersion diagram. As shown in the Fig. \ref{fig: Sweep}(a),
by increasing $C_{\mathrm{0}}$, the dispersion diagram shifts downwards
and consequently the EPDs move in the same direction. In the next
step, we study the effect of the modulation depth $\delta_{\mathrm{c}}$
perturbation on the dispersion diagram in Fig. \ref{fig: Sweep}(b).
By increasing the modulation depth, the band-gaps stretch out and
become wider, meaning that EPDs at both edges of one band-gap move
further apart from each other in frequency. As the third parameter,
we explore the temporal modulation frequency $\omega_{\mathrm{m}}$
variation on the location of the band-gaps and EPDs. Fig. \ref{fig: Sweep}(c)
exhibits a similar trend of changes compared to those in Fig. \ref{fig: Sweep}(a).
Finally, we examine the variation of spatial modulation frequency,
$\beta_{\mathrm{m}}$, shown in Fig. \ref{fig: Sweep}(d). It is seen
from this figure that a different behavior is obtained compared to
varying the previous parameters. Here, by increasing the spatial modulation
frequency $\beta_{\mathrm{m}}$, band-gaps become wider and move toward
higher frequency in the dispersion diagram; thus, EPDs move to higher
frequencies as well.

As discussed in the Introduction and as it is clear from the plots
in Fig. \ref{fig: Sweep}, the eigenvalues at EPDs are exceedingly
sensitive to perturbations of parameters of a time varying system
\cite{Wiersig2016Sensors,Kazemi2019Exceptional,Kazemi2019Experimental,Kazemi2019Ultra}.
Here we show that the sensitivity of a system\textquoteright s observable
to a specific variation of a parameter is boosted due to the degeneracy
of eigenmodes. As an example, to assess the degree of perturbation
of wavenumbers, we consider the first EPD and the first band-gap with
the negative real part of wavenumber (indicated by a gray circle in
Fig. \ref{fig: Dispersion}). We define the relative system perturbation
$\Delta$ as

\begin{equation}
\Delta=\frac{X_{\mathrm{pert}}-X_{\mathrm{EPD}}}{X_{\mathrm{EPD}}},\label{eq: perturbation_touchstone}
\end{equation}
\begin{figure}[t]
\begin{centering}
\includegraphics[width=2.7in]{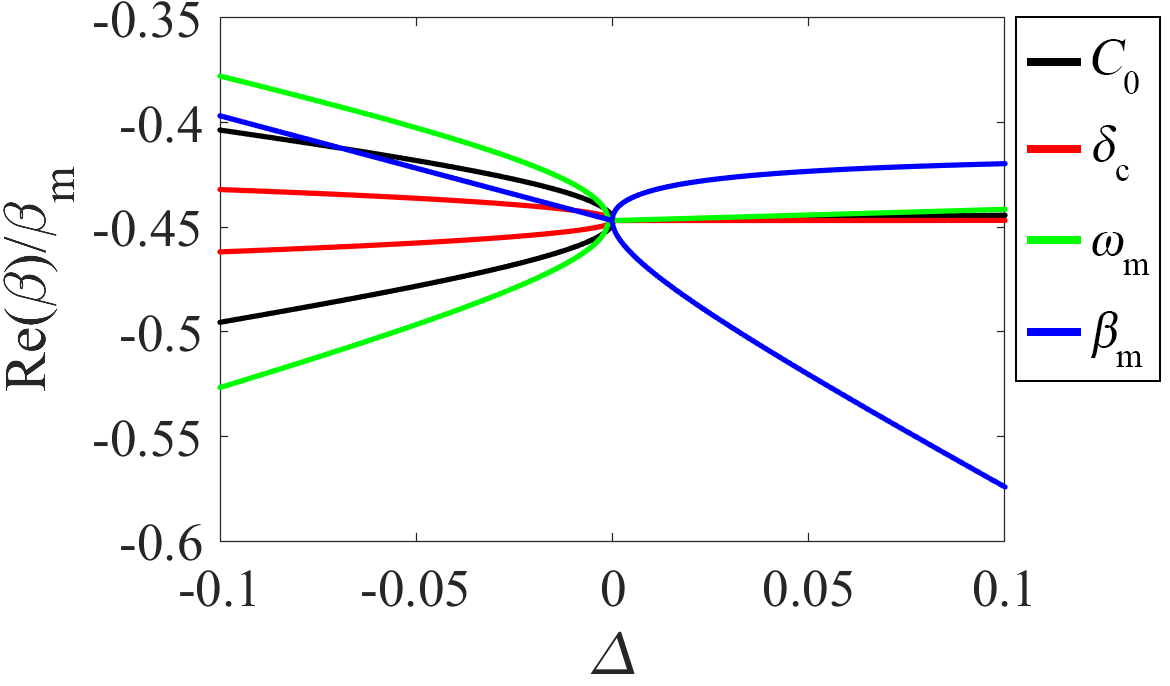}
\par\end{centering}
\caption{Sensitivity of the real part of the propagation wavenumber to a parameter
perturbation (one at the time) around the EPD.}

\label{fig: Perturbation}
\end{figure}
where $X_{\mathrm{EPD}}$ is the unperturbed parameter value that
provides the EPD condition, and $X_{\mathrm{pert}}$ is its perturbation.
We consider variations of $C_{\mathrm{0}}$, $\delta_{\mathrm{c}}$,
$\omega_{\mathrm{m}}$, and $\beta_{\mathrm{m}}$, one at the time.
The calculated real part of the wavenumber near the first EPD at $\omega/\omega_{\mathrm{m}}=3.91$
is shown in Fig. \ref{fig: Perturbation}. We conclude from the extracted
results that the individual variation of the parameters of $C_{\mathrm{0}}$,
$\delta_{\mathrm{c}}$, and $\omega_{\mathrm{m}}$, show similar sensitivity
behavior, i.e., the real part of the wavenumber splits for $\Delta<0$.
In contrast, variation of $\beta_{\mathrm{m}}$ has an opposite effect
on the dispersion diagram, i.e., the real part of the wavenumber splits
for $\Delta>0$. Note that the $\omega_{\mathrm{m}}$ perturbation
shows the highest sensitivity. Higher sensitivity is obtained when
the bifurcation of the dispersion diagram is wider. Furthermore, the
$\beta_{\mathrm{m}}$ perturbation response shows an opposite trend
to that of the other three parameters.

We explain the extreme sensitivity by resorting to the general theory
of EPDs. Note that a perturbation in $\Delta$ value leads to a perturbed
matrix $\mathbf{\underline{T}}(\Delta)$. Consequently, the two degenerate
eigenvalues occurring at the EPD change considerably due to a small
perturbation in $\Delta$, resulting in two distinct eigenvalues $\beta_{\mathrm{\mathit{p}}}(\Delta)$,
with $p=1,2$, close to the first EPD. The two perturbed eigenvalues
near an EPD are represented by a single convergent Puiseux series
(also called fractional power expansion) where the coefficients are
calculated using the explicit recursive formulas given in \cite{Welters2011Explicit}.
An approximation of $\beta_{p}(\Delta)$ around a second-order EPD
is given by

\begin{equation}
\beta_{p}(\Delta)\approx\beta_{EPD}+(-1)^{p}\alpha_{1}\sqrt{\Delta}.\label{eq:Puiseux}
\end{equation}

\begin{figure}
\begin{centering}
\includegraphics[width=2.7in]{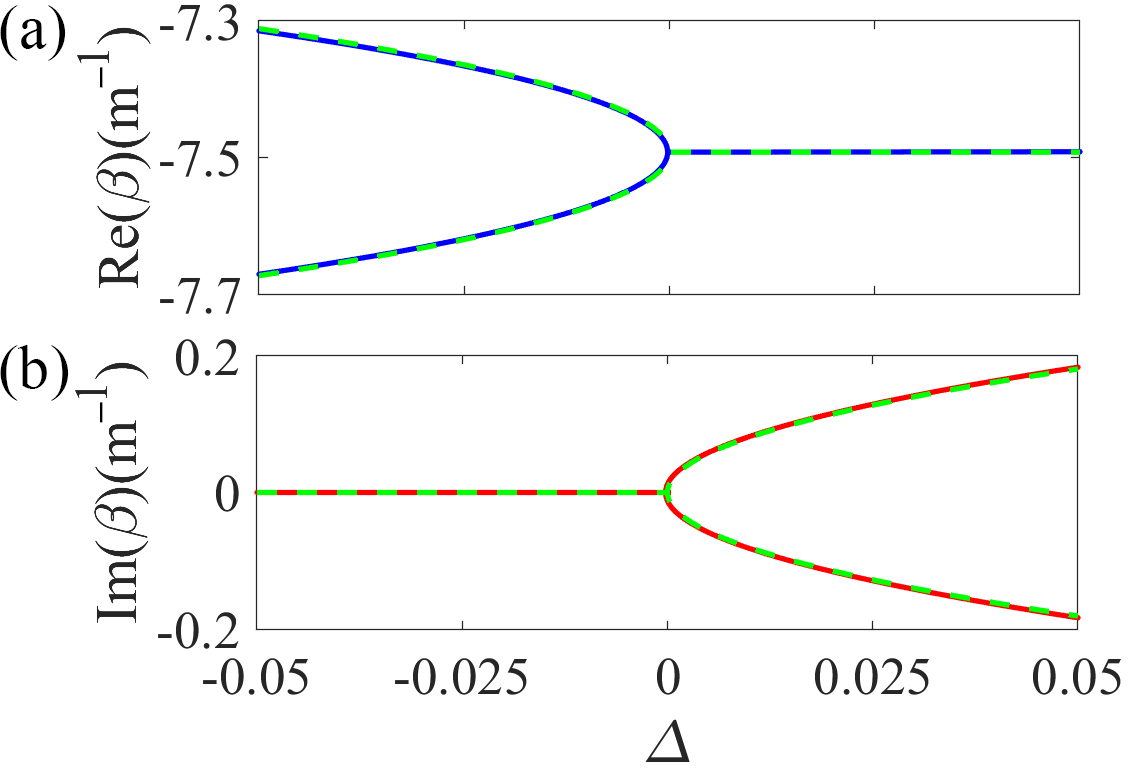}
\par\end{centering}
\caption{The Puiseux fractional power series expansion in Eq. (\ref{eq:Puiseux})
(green dashed lines) describes the bifurcation of the real and imaginary
parts of the two wavenumbers when a system parameter ($\delta_{\mathrm{c}}$
in this case) is perturbed. The Puiseux series result is in excellent
agreement with the wavenumbers evaluated using Eq. (\ref{eq: T_mat})
(blue and red solid lines).}

\label{fig: Puiseux}
\end{figure}

Following \cite{Moro1997Lidskii,Welters2011Explicit,Hanson2018Exceptional},
we calculate $\alpha_{1}$ as

\begin{equation}
\alpha_{1}=\sqrt{\left(-\frac{\frac{\partial H}{\partial\Delta}(\Delta,\beta)}{\frac{1}{2!}\frac{\partial^{2}H}{\partial\beta^{2}}(\Delta,\beta)}\right)},\label{eq:PuiseuxCoeff}
\end{equation}
evaluated at the EPD, i.e., at $\Delta=0$ and $\beta=\beta_{\mathrm{EPD}}$,
where $H(\Delta,\beta)=\mathrm{det}[\underline{\mathbf{T}}(\Delta)-\beta\underline{\mathbf{I}}]$.
Equation (\ref{eq:Puiseux}) indicates that for a small perturbation
$\Delta\ll1$ the eigenvalues change dramatically from their original
degenerate value due to the square root function. As an indicative
example, we consider the single STM-TL with parameters as used in
Fig. \ref{fig: Dispersion}. In this example, we select the EPD indicated
by the gray circle in Fig. \ref{fig: Dispersion}(a) with $\omega/\omega_{\mathrm{m}}=3.91$,
and $\beta_{\mathrm{EPD}}=-7.49\,\mathrm{m}^{-1}$ as the unperturbed
EPD operation point. In this example the perturbation parameter is
the modulation depth, $\Delta=(\delta_{\mathrm{c}}-\delta_{\mathrm{c,EPD}})/\delta_{\mathrm{c,EPD}},$
and the Puiseux series coefficients is calculated as $\alpha_{1}=j0.81\,\mathrm{m}^{-1}$.
The result in Fig. \ref{fig: Puiseux} exhibits the two branches of
the exact perturbed eigenvalues $\beta$ obtained from the eigenvalue
problem in Eq. \ref{eq: T_mat} when the system perturbation $\Delta$
is applied. Moreover, this figure shows that such perturbed eigenvalues
can be estimated with very good accuracy by employing the Puiseux
series (green dashed lines) truncated at its first order. For a positive
but small value of $\Delta$, the imaginary part of the eigenvalues
experience a sharp change, while its real part remains constant. Moreover,
a very small negative value of $\Delta$ causes a rapid variation
in the real part of the eigenvalues. This feature is actually one
of the most extraordinary physical properties associated with the
EPD concept, and it can be exploited for designing ultra-sensitive
sensors \cite{Chen2017Exceptional,Hodaei2017Enhanced}. This kind
of STM-TL with capacitance variation is feasible within the realm
of current fabrication technologies. Varactor-loaded transmission-line
could be a proper alternative for implementing this kind of structure
\cite{Ellinger2003Varactor}. In addition, in recent years several
tunable materials have been employed to conceive tunable devices,
such as graphene \cite{Rouhi2019Multi} and liquid crystal \cite{Komar2018Dynamic},
and they are a promising candidate for realizing spatiotemporally
varying TLs.

\section{Conclusion}

A single STM-TL supports EPDs of second order directly induced by
spatiotemporal modulation of the distributed (per-unit-length) capacitance.
For its occurrence, an EPD does not need the presence of time-invariant
gain or loss elements, as in PT symmetry, and it does not need two
coupled transmission lines either. Here space and time modulation
are not used to generate nonreciprocity or to enhance EPD properties
but rather as a direct way to generate EPDs. This is in analogy to
what was shown in \cite{Kazemi2019Exceptional} where time modulation
was used to directly induce EPDs in a \textit{single} resonator, without
the need to resorting to two couple resonators with loss and gain
as implied by PT-symmetry \cite{Schindler2011Experimental}. We have
investigated how to perturb an EPD condition by slightly perturbing
system parameters, and how this perturbation modifies the degenerate
eigenvalues. We have shown that small changes in a TL constitutive
parameters lead to a very strong variation of the TL wavenumber and
how this is predicted by the Puiseux fractional expansion series,
suggesting a novel approach to design extremely sensitive sensors
based on waveguide propagation.

\section{Acknowledgment}

This material is based upon work supported by the National Science
Foundation under Grant No. ECCS-1711975.

\bibliographystyle{IEEEtran}

\end{document}